# Navigation Framework for Blind and Visually Impaired Persons based on Sensor Fusion

Chathurika S. Silva
Faculty of Technology, University of Colombo,
Sri Lanka
chathurika@iat.cmb.ac.lk

Prasad Wimalaratne
University of Colombo School of Computing
Sri Lanka
spw@ucsc.cmb.ac.lk

*Abstract*—Individuals who are differently-able in vision cannot proceed with their day-to-day activities as smoothly as other people do. Especially independent walking is a hard target to achieve with their visual impairment. Assistive electronic travel aids equipped with different types of sensors are designed for visually impaired persons to assist their safe navigation. The amount of research on combining multiple sensors in assistive navigation aids for visually impaired navigation is limited. Most work is targeted at sensor integration but not at sensor fusion. This paper aims to address how sensor fusion and integration will be used to improve the sub-processes of visually impaired navigation and the way to evaluate the sensor fusion-based approach for visually impaired navigation which consists of several contributions to field sensor fusion in visually impaired navigation such as a novel homogeneous sensor fusion algorithm based on extended Kalman filter, a novel heterogeneous sensor integration approach, and a complementary sensor fusion algorithm based on error state extended Kaman filter. Overall this research presents a novel navigational framework to integrate obstacle detection, obstacle recognition, localization, motion planning, and current context awareness with sensor fusion.

## I. INTRODUCTION

Individuals who are differently-able in vision find it challenging to carry out day-to-day activities such as independent walking compared to others. Especially independent walking is a hard target to achieve with their visual impairment. Assistive technology to aid the mobility of blind people is an emerging area where several scientific contributions have been made to assist the navigation of visually impaired people by mainly facilitating the autonomous execution of intelligent environments and accessible context-aware smart navigation aids.

However, most assistive navigation aids depend on the measurements acquired by a single type of sensor attached to the user. The amount of research on combining multiple sensors in assistive navigation aids for visually impaired navigation is limited. Another observation was a lack of integration of navigational sub-processes such as obstacle detection, localization, and motion planning among the navigational aids for visually impaired persons in the literature.

This research investigates and develops a navigation framework with sensor fusion-based obstacle detection, recognition, localization, and motion planning. However, exploring the appropriate type and number of sensorial channels of complementary sensors to aid visually impaired navigation is a critical challenge [1].

## II. BACKGROUND

Blind navigation is a cognitively demanding task since the blind person does not have access to contextual information and spatial orientation, requiring moment-to-moment problem-solving [2]. A guide dog and a white cane are traditionally used as primary navigation aids for visually impaired persons [3]. A guide dog is costly and has a limited lifetime, whereas the white cane can only sense his environment's immediate surroundings. Therefore, researchers have attempted to develop various secondary travel aids for visually impaired navigation over the past decades in addition to the white cane. As a result, travel aids with different sensors and technologies have been investigated.

### A. Sub-Processes of Navigation

Mobility consists of several sub-processes such as obstacle detection, obstacle recognition, localization, motion planning, etc. [4]. Static and dynamic obstacles that can collide with the navigator are considered under obstacle detection. Obstacle recognition can ensure the existence of the obstacle while giving additional information on the detected obstacle. Most of the navigation aids reported in the literature consist of single sensor-based object detection [5] [6], recognition [7], localization [8], and motion planning [7]. Therefore, most of the research has stated that it was difficult to gain the optimum outcome of those sub-processes due to the distortions of sensor data. It was also mentioned in the literature [9] that the performance of the navigation sub-processes can be improved through sensor fusion.

### B. Sensor Fusion and Integration

According to Luo, two fundamental approaches combine information from different sensors: Integration and Fusion.
1. Integration: using information from multiple sensors to assist in the systems goal achievement and allow the data from each sensor to function as an independent input to the system controller [10]
2. Fusion: information from different sensors is combined to take them to one standard representational format. Information may belong to multiple sensors during a single period or from a single sensory device over an extended period [10].

Sensor fusion methods and techniques can be classified based on levels of abstraction, the relation of data sources, and the type of sensors as shown in Fig. 1 in our previous publication [11].

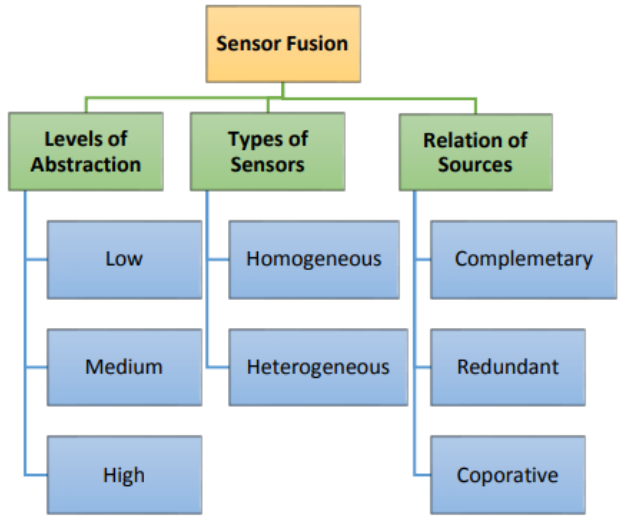

**Fig. 1.** Sensor Fusion Taxonomy [11]

Sensor fusion algorithms combine sensory data to reduce uncertainty in environment perception and help to make timely and situation-appropriate decisions. A strong fusion algorithm gives preference to some data points over others depending on the pros and cons of sensors relevant to a specific use case.

*C. Evaluation of Visually Impaired Navigation*

Evaluation of sensor-based assistive navigation aids for visually impaired persons can be broadly classified into two: real-world environment-based evaluations and simulated environment-based evaluations. Real-world environment-based evaluations can be conducted in both indoor and outdoor environments. GPS sensor-based assistive aids are common for the outdoor navigation of visually impaired people [12] [13] [8]. Indoor navigation aids remain more challenging and primarily based on sonar, vision, IR, laser etc. [14]- [15]. The selection of subjects for the evaluations conducted in the real-world environment is also a significant concern in visually impaired navigation. Only blindfolded persons participated in some studies [16], while some other studies consist of blind and blindfolded subjects. Most of the evaluations consist of males and females and subjects with different age levels. Experiments in the real world are primarily conducted in controlled environments. Especially indoor environments are arranged with known obstacles and landmarks [17]. Pre-located routes, sidewalks and less crowded lanes are selected as the experimental setups for the evaluations conducted in outdoor environments [18]. Real-world evaluations of sensor fusion-based sub-processes such as obstacle detection [19], obstacle recognition [20] [21], map matching, and localization [22] in visually impaired navigation are minimum.

However, conducting evaluation experiments with assistive navigation in real-world environments with visually impaired subjects is challenging due to several factors. The evaluation of assistive navigation in the real-world setting can pose safety risks to the subjects of the experiments and may also cause disturbance to the other pedestrians. Setting up controlled environments is expensive and usually requires approvals from several parties [23]. Hence simulation-based usability evaluation experiments are a pragmatic and cost-effective approach in such studies. In literature, most visually impaired navigation simulations use virtual reality to navigate in a virtual environment using non-visual information [24] [25].

The purpose of this paper is to investigate sensor fusion and integration approaches used to complement the limitations of different sensor types used in visually impaired navigation.

III. METHODOLOGY

The main contribution of this paper is the development of a framework for the navigation of visually impaired people. In this framework mainly five components are identified through literature obstacle detection, obstacle recognition, localization, and motion planning. All these five components are sub-processes of visually impaired navigation. Therefore the identified sub-processes of visually impaired navigation are included as components of the proposed framework.

All the components of the proposed framework are experimentally investigated, designed and evaluated. The research challenge included individually designing and developing the components and subsequently integrating them into the framework. So that, they work as a whole to achieve the purpose of navigation.

The proposed work adopts a constructive research methodology [26] based approach. Constructive research intends to produce 'novel constructs', such as models, frameworks, and prototypes, to solve the identified problems. Thus, constructive research is considered appropriate to narrow the gap between research and practice [27].

The high-level overview of the proposed framework is shown in Fig. 2. According to this figure inputs of the walking environments are obtained via proprioceptive and exteroceptive sensors and personal factors of individuals are obtained via a smartphone application. Navigation processes which are included as components of the framework such as obstacle detection, obstacle recognition, localization, and motion planning are based on the inputs obtained via sensors. Finally, the outputs of the framework components are sent as audio and tactile feedback to visually impaired persons.

**Walking Environment:** Consists of all the static, dynamic obstacles and other variables that affect visually impaired persons' navigation.

**Personalized Smartphone Application:** This provides the personal factors such as age, gender, height and visual status of the visually impaired person to customize the navigation prototype further.

**Sensors:** Sensors are used to measure or detect a property of the environment. Exteroceptive sensors are used to perceive the surrounding environment correctly. E.g., Camera, Ultrasonic sensor. Proprioceptive sensors are the sensors that sense ego properties. E.g., GPS, Inertial Measurement Unit (IMU).

**Environment Perception:** In navigation, the visually impaired user has to detect objects and events that will immediately affect the navigation and recognize possible obstacles. Identifying the obstacles that are already detected will ease the recognition process while saving computational processing power.

**Localization:** Localization is the method to determine the position and orientation within the environment.

**Motion Planning:** Motion planning based on the output of localization and Environment perception sub-components. The output of motion planning is a planned path from start to goal.

**Feedback:** Finally, multimodal feedback is given to blind persons. Different feedback modes give feedback for obstacle detection, recognition, and motion planning.

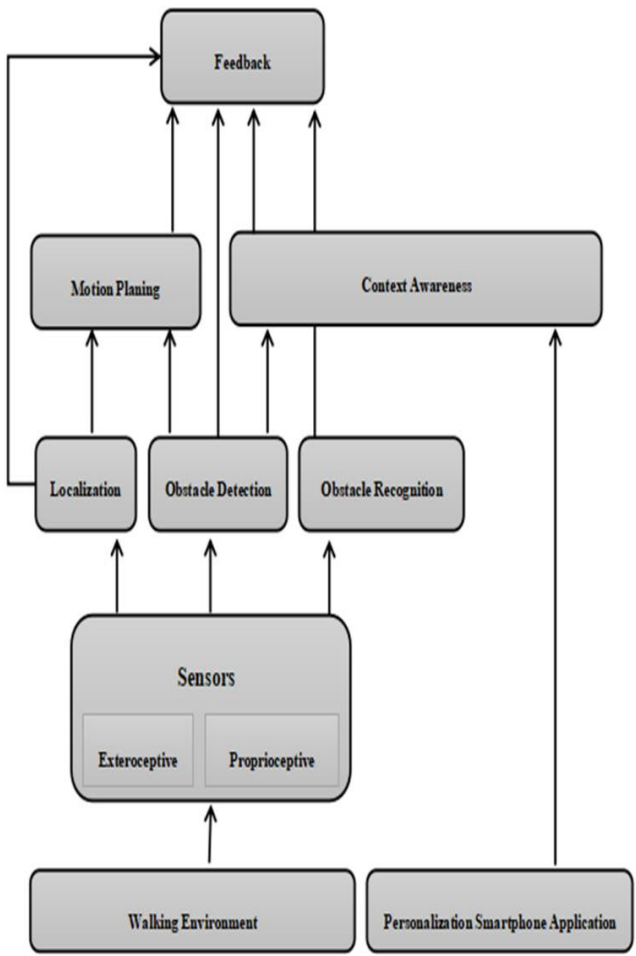

**Fig.2.** High-level Overview of the Framework

*D. Obstacle Detection and Recognition*

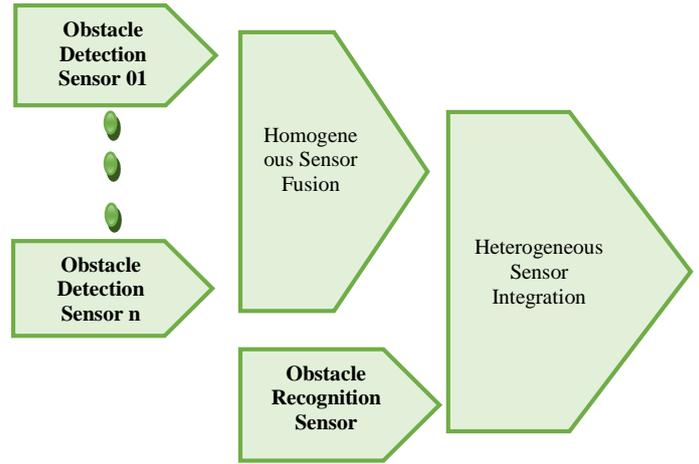

**Fig. 3.** Abstract View of the obstacle detection and recognition

Obstacle detection alarms on possible threats which can be hazards for visually impaired navigation and obstacle recognition can provide more detailed information about the detected obstacles. However, both obstacle detection and recognition can provide false alarms about objects which can occur due to the limitation of the sensors. Therefore, as shown in Fig. 3, the limitations of obstacle detection sensors are reduced through the homogeneous sensor fusion approach, and the heterogeneous sensor integration approach is used to enhance obstacle recognition via obstacle detection and obstacle recognition sensors.

The Sonar sensor is selected as the obstacle detection sensor. Fusion of sonar sensors yields better results since uncertainties occur due to the wide beam width of sonar sensors and environmental influences such as air temperature, humidity, air pressure and air currents. EKF fuse sonar sensor signals. A vision sensor is selected to complement with sonar sensor since the vision sensor can perform both obstacle detection and recognition. A heterogeneous sensor integration approach is used to fuse sonar and vision sensors since sonar and vision sensors have asynchronous sampling rates in data processing where the processing speed of the sonar sensor is higher than that of the vision sensor. Therefore, sonar speeds the process of visual data by selecting only the regions which have an obstacle. The vision sensor is triggered whenever the sonar sensor detects an obstacle and processes that video frame only.

Two inclined sonar sensors attached to the waist belt are used to detect ground-level obstacles rather than using a single inclined sensor. EKF is selected as the most suitable sensor fusion approach to fuse two sonar sensors. EKF is an extension of the linear Kalman filter. The Kalman filter-based sensor fusion can provide a closer approximation to the true state of the system than the single sensor alone [28].

## IV. IMPLEMENTATION

The wearable belt shown in Fig. 4 with SRF 05 ultrasonic sensors are used to detect the obstacles. A microcontroller processing module processes sonar signal data in the Arduino UNO board based on ATmega 328. Google cloud-based image processing uses a label detection algorithm for obstacle recognition. The use of cloud computing-based processing overcomes the limitations in the computational power of mobile and embedded devices. The camera is triggered to obtain the video frames to identify the properties of the detected objects. The U-Blox NEO 7M GNSS receiver is chosen as the GPS sensor. The GPS provides a certain amount of data such as position, speed over the ground, time, etc. The IMU and GPS are connected to an Arduino Uno microcontroller. Raspberry PI 3B+ is chosen to record and fuse the data. Since the Raspberry PI 3B+ has four cores operating at 700 MHz, it should have enough computational power to run real-time sensor fusion. The system uses coin vibration motors to generate tactile feedback. The wearable sensor belt consists of five-coin vibration motors. Sensors are placed outside of the belt, where they face the surrounding environment, and tactile units are attached to the belt such that they are in contact with the body (around the waist). Audio feedback is provided via an earphone provided to one ear allowing the other ear to open to the background noise.

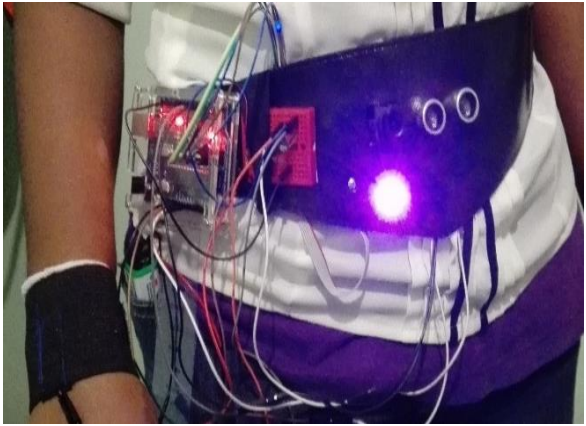

**Fig. 4.** Proof of Concept of Prototype

## V. EVALUATION

The experimental protocol for evaluating sub-processes of the framework involves selecting subjects, setting up the controlled environments, and conducting the experiments. Subsequently, evaluation experiments are carried out in a controlled, real-world environment. The results are then analyzed, and conclusions are drawn under each sub-processes of obstacle detection, obstacle recognition, localization, and motion planning.

*A. Experimental Protocol*

These research subjects were ten users of both genders (four female and six male users) and different age levels (eight young users around 22-37 years and two old users around 70 years old). Out of them, five subjects are blindfolded, and their visual status is assumed to be as blind. The other three subjects have visual impairments, such as refractive error and age-related visual losses, and two subjects are blind. All the participants had normal hearing abilities, confirming that they do not have any other disabilities [32].

During the experiment, both physical well-being and privacy were well protected. Safety measures were taken to minimize risks such as falls during the experiment. Extensive training was given to all the study participants to maintain the consistency of the results between them. Close attention was paid to the comfort of the subjects allowing them to pause the experiment any time they felt tired. The navigation environment included an area with different obstacles. This evaluation environment was adjusted beforehand and was not seen by the participants before participating in the evaluations.

*B. Evaluation of Obstacle Detection*

The average rate for left-side obstacle detection was about 89%, right-side obstacle detection was about 86%, and front obstacle detection was 98%. Drop-off detection has the lowest detection accuracy, which is 67%. When there is more than one obstacle around the user, he/she gets confused in reacting to the multiple vibration feedback.

A single sonar sensor measurement consists of many uncertainties, as shown in Fig. 5. Therefore, it is proposed to obtain more reliable sonar measurements with less uncertainty using the homogeneous fusion of several sonar sensors.

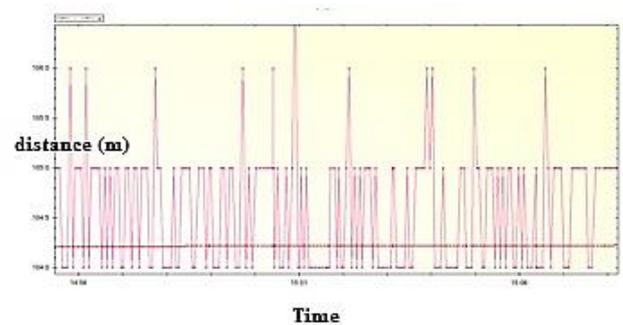

**Fig. 5.** Measurement Uncertainties in Single Sonar Sensor

The fusion of two ultrasonic sensors using EKF has been considered, and both sensors are supposed to measure the distance between the visually impaired person and the obstacle. Therefore, the state vector comprises two distance measurements where both sensors contribute equally to the state estimation.

The SRF 05 is the ultrasonic sensor module used to detect the POC prototype's obstacle detection. The average noise of the SRF 05 ultrasonic sensor is 0.3 cm, according to the specifications of the datasheet [123]. Therefore, the ultrasonic reading standard deviation is 0.3, making the variance 0.3 x 0.3 = 0.09. This gives the covariance matrix of the measurement noise (R) in the matrix 1.

$$R = \begin{bmatrix} 0.09 & 0 \\ 0 & 0.09 \end{bmatrix} \quad (1)$$

All the diagonal elements are equal to the variance of sensor noise. All the off-diagonal elements are zero-based on the assumption that the noise of one sonar sensor does not affect the noise of the other sonar sensor.

Following is the covariance of process noise (Q) which represented by matrix 2.

$$Q = \begin{bmatrix} 0.001 & 0 \\ 0 & 0 \end{bmatrix} \quad (2)$$

The EKF algorithm is evaluated based on the real-time distance measurements (observations) from two ultrasonic sensors. The first observation is the initial value (base case) for the state estimates. The initial value for the prediction estimate is set to 1.

The green line in Fig. 6 illustrates the fusion outcome, whereas the red and blue lines represent the raw data from the two ultrasonic sensors.

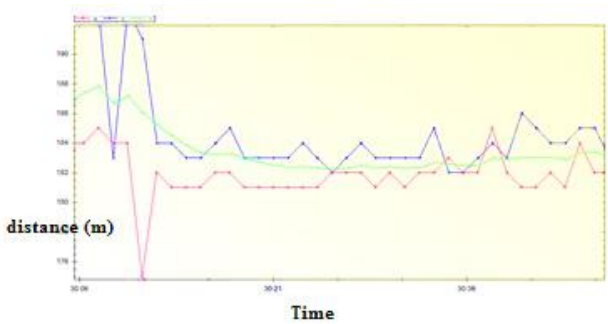

**Fig. 6.** Homogeneous Fusion of Two Ultrasonic Sensors

*C. Obstacle Recognition*

The average feedback period between sending and receiving the captured image was determined for pictures with different resolutions. The feedback times corresponding to different resolutions are shown in Fig. 7.

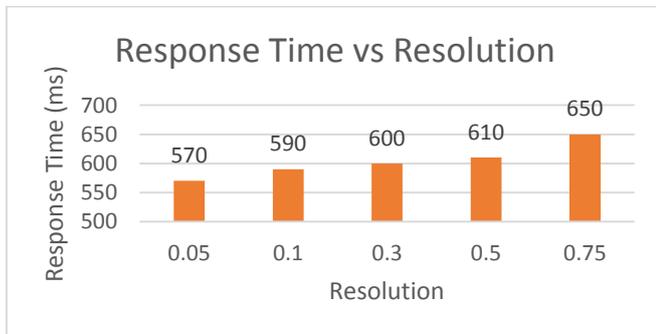

**Fig. 7.** Change in Response Time with Resolution

The lower resolutions obtained by compressing the same set of frames had shorter response times than the default resolution.

*D. Evaluations of Localization*

Inertial measurements from IMUs with position measurements from a GPS produce an accurate localization estimate for a visually impaired navigator. Both of these sensors use different measurement methods that are unlikely to fail for the same reason and are complementary, which can be used together in practice to minimize the limitations of each other. Over time, the drift accumulated by IMU can be compensated by the bounded error positioning updates provided by the GPS.

*i.    Evaluations of GPS-based Localization*

An experiment is done by stopping at four different locations and standing still there for a few seconds at each location during 230 m walks. The GPS measurements (WGS84 coordinates) are recorded at each of these locations separately. Then the recorded coordinates convert to geocentric coordinates and are plotted in Fig. 8 using different colours for each location. The GPS waypoints cluster with different colors, showing the variability in GPS measurements even when the walker is standing still at a particular location.

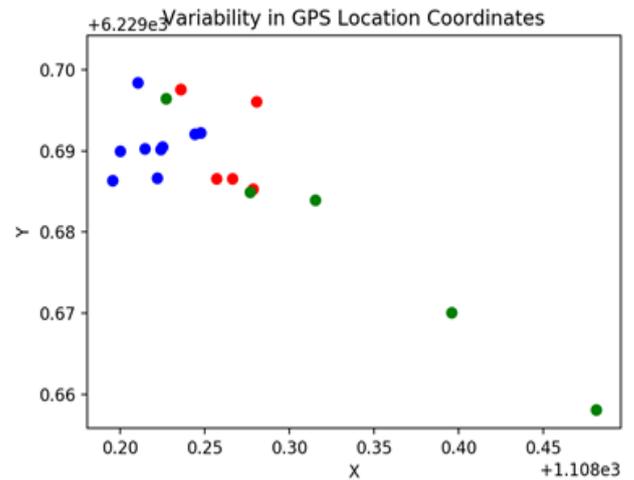

**Fig. 8.** GPS Waypoint Clusters with Noise

Ground truth is obtained via GPS waypoints. However, GPS waypoint data can be noisy and may have a low sampling rate, as shown in Fig. 9. Noise is caused by high-rise buildings or poor GPS signal reception. Noisy waypoints can generate erroneous ground truth trajectories. Therefore, calibration of raw GPS data is essential in navigation. However, there is no ground truth data to validate the calibrated GPS data in practice. It is believed that human-labelled data can be almost 100% accurate, and it is widely used to explore ground-truth datasets to evaluate map-matching algorithms [33]. A human labelling process involves both cognitive work and manual work. Since this process involves too much human intelligence and action, it is usually not feasible to apply pure human labelling to large GPS datasets.

In this work, a manual human labelling approach is undertaken to alleviate erroneous ground truth trajectory since the GPS data set is small. Therefore, the method of combining GPS waypoints selected from consecutive clusters that are spatially close to each other is shown in Fig. 9.

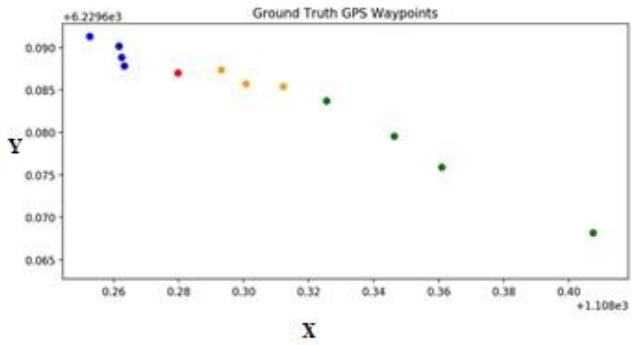

**Fig.9.** GPS Waypoints Selected as Ground-Truth

### ii. Inertial Sensor-based Navigation

A stand-alone GNSS does not offer the precision and accuracy required for the localization of visually impaired persons. A commonly used approach to improve GNSS accuracy is to employ additional sensors to improve localization through sensor fusion. Sensors often used for this purpose are inertial sensors.

The IMU was placed tightly onto the chest using strap bands. The direction of the tri-axial sensors was when standing up straight, the x-axis pointing forward in the direction of the nose, the y-axis to the right, and the z-axis down.

*IMU Calibration-* Sensor calibration is essential for combining different sensors to estimate the navigation state. Primarily, sensors are calibrated during their manufacturing process, and the specification sheet relevant to each sensor reflects the corresponding calibrated parameters. However, certain calibration steps are undertaken for some sensors before putting the sensors into operation.

There is often a small error in the average signal output, even when there is no movement. This is what is also known as Sensor Bias. Therefore, the MPU-6050 needs to be calibrated before being used for the first time. The calibration code, which is done by Ródenas [34], is used in this work. This error can be removed by applying an offset to the raw accelerometer and gyroscope sensor readings. The offset needs to be adjusted until the gyroscope readings are zero (no rotation), and the accelerometer records the acceleration due to gravity pointing directly downwards. Before starting the calibration process, the MPU6050 module is placed in a flat and level position. Moreover, any character is sent to the serial monitor. The accelerometer and gyro offsets are set to zero at the beginning of the process. Then calculate the mean value for each offset by taking 1000 readings from the IMU. These values are then entered into the IMU to become the new offsets. The calibration routine continues to take the mean IMU readings until the calibration is within a certain tolerance.

*Evaluations of ES-EKF-based Localization-* Before each field test, a calibration routine was performed. Then, data was recorded when each tester performed walking forward. Mainly two assumptions are made when fusing the IMU and GPS measurements: 1. There is no delay between IMU measurement and frame construction such that IMU packet creation time is negligible. 2. IMU receives Universal Coordinate Time (UTC) from Arduino, and GNSS receives UTC based on atomic clocks onboard satellites.

The accelerations and rotational rates from the IMU are integrated with the motion model to calculate position, velocity, and orientation. The GPS measurements are incorporated at a much slower rate (once a second) whenever available. The GPS measurements are used to correct the predicted state.

The validation experiment compares the estimated and ground-truth walking path. First, an area to be walked is marked with some key points, and it was approximately 110 m long. Second, the average recorded walking speed was 1.52 m/s, with a total walking time of 81 seconds. Finally, upon completing the experiment, the presented estimated navigation output of the ES-EKF algorithm is plotted, and the recorded ground truth is overlaid.

Ground truth location data is plotted against the estimated location data given by the ES-EKF algorithm within the same axis, as shown in Fig. 10.

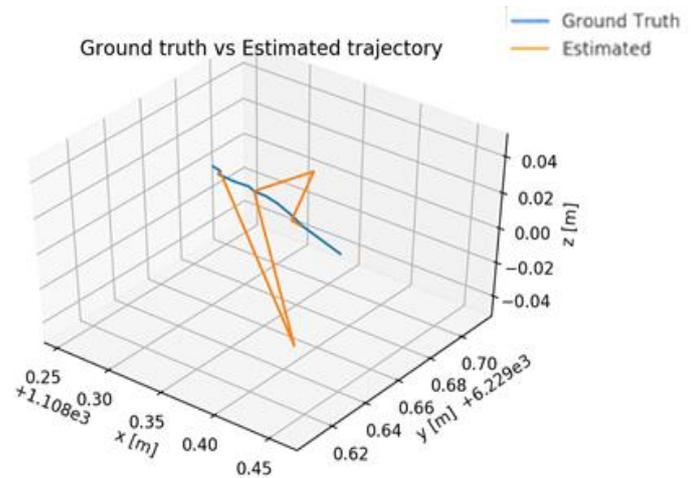

Fig.10. Ground Truth vs Estimated Trajectories (Using Raw Data)

There are two ways to extract useful data from the IMU. One way is to read the raw sensor data values and compute the new orientation. Fig. 10 shows the estimated trajectory calculated using the raw values. The second method is to pull the data out of the MPU's onboard DMP. The DMP offload processing typically has to take place on the microprocessor. It maintains an internal buffer that combines data from the gyro and accelerometer and computes orientation. The DMP also takes care of applying the offsets. Fig. 11 shows the estimated path with the DMP data.

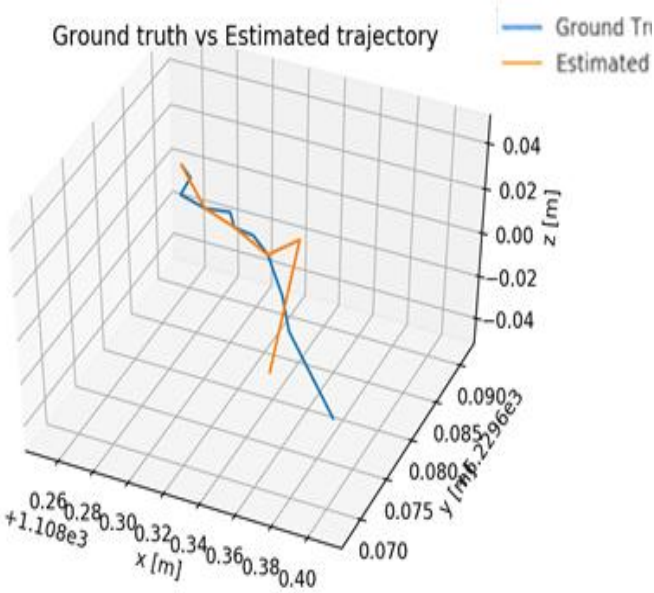

**Fig. 11.** Ground Truth vs Estimated Trajectories (Using DMP Data)

The benchmarking of estimated localization data obtained from ES-EKF-based fusion of GPS and IMU with the ground truth reference shows only a 28.37% relative error percentage.

Further, the mean total distance error for the points estimated is 3.3 m, in the raw data case with a peak distance error of 12.75 m compared with the ground truth. In the DMP case, the mean error was reduced to 1.74 m with a peak error of 6.22 m, respectively. This proves that the use of data from the DMP improves the estimated path given by the ES-EKF algorithm rather than using the raw data directly.

Although many messages could be generated, message flooding causes severe latency for user feedback, and the user may quickly get confused and annoyed. Therefore, the system should provide only the most important messages that suit the user's particular need in an ordered sequence.

The distances to the obstacles detected by left, right, and front sonar sensors are given via tactile feedback. The distances to the closest object are difficult to deliver using audio messages, as it keeps changing from time to time. Therefore, tactile feedback is defined, and the intensity of the vibration changes following the distance to the closest object, i.e., if the object appears from a far distance, the frequency of the vibration is less, and if it appears very closely, the intensity gradually increases.

*E. Comparison with the Recent Literature*

Table 3 shows the quantitative comparison of the results obtained from this study with the recent literature which are remotely/logically equivalent to the proposed approach.

Following comparison of current work to the state-of-art proves the novelty and the strength of the proposed approach.

Table. 3. Comparison of Navigation Sub-processes with the Recent Literature.

| Navigational Sub-processes | Proposed Approach | Literature |
|---|---|---|
| **Obstacle Detection** | Left-side obstacle detection 89%, right-side obstacle detection 86%, and front obstacle detection 98%. Drop-off detection has the lowest detection accuracy, which is 67%. | For the case of the front obstacle, the capacity of avoidance is high while left and right objects can reach from 45 % to around 62 % of avoidance capacity. [9] |
| **Obstacle Recognition** | The average response time was given as 604ms. | The average response time given for different resolutions of the images was 580ms. [35] |
| **Localization** | Estimated localization data obtained from ES-EKF-based fusion of GPS and IMU sensors shows a 28.37% relative error percentage respective to the ground truth obtained from GPS. | The proposed system can reliably provide precise locations information with a median error of approximately 0.27 m. [36] |
| **Motion Planning** | A mean total distance error of 1.74m was recorded for the walking path of 110m. | Approximately 2.75m average distance error is recorded for 100m distance. [37] |

VI. CONCLUSION AND FUTUREWORKS

This research is motivated by the observation that there is a lack of integration of navigational sub-processes such as obstacle detection, localization, and motion planning among the navigational aids for blind and visually impaired persons. Literature in the recent past showed that the existing navigational aids designed for the above purpose consist of a single type of sensor, limiting navigation and localization accuracy and efficiency. Thus, this research proposes a navigational framework with several components that represent navigational sub-processes with homogeneous and heterogeneous sensor fusion. This research presents a novel framework to integrate obstacle detection, obstacle recognition, localization, and motion planning with sensor fusion. The proposed approach consists of several contributions to field sensor fusion in visually impaired navigation—for instance, a novel homogeneous sensor fusion algorithm based on EKF to fuse multiple sonar sensors. Subsequently, a novel heterogeneous sensor integration approach is proposed to integrate vision and sonar sensors. Moreover, a complementary sensor fusion algorithm based on ES-EKF is introduced to fuse inertial and GPS sensors for localization. The proposed framework can be extended to include more navigational sub-processes with additional sensors to provide an independent navigation experience for visually impaired people, and the proof-of-concept implementations are scalable to incorporate the extensions of future assistive technologies.

In future work, the walking behaviour of visually impaired persons can be modified according to the motion predictions of dynamic obstacles. Motion prediction attempts to estimate the future positions, headings, and velocities of all dynamic objects in the environment over some finite horizon. This is important for the motion planning problem, as it allows us to plan future actions for visually impaired navigation based on the expected motions of other dynamic objects. The predicted paths also enable making sure that the path of visually impaired persons does not collide with any future objects at a future time. In future work, expanding the framework by adding more components via validated constructs and performance optimization of the overall framework can be performed.

**Disclosure statement:** The authors report there are no competing interests to declare.

**Conflict of Interest:** There are no conflicts of interest. All procedures performed for studies involving human participants were in accordance with the ethical standards stated in the 1964 Declaration of Helsinki and its later amendments or comparable ethical standards. Informed consent was obtained from all participants.

REFERENCEES

**Chathurika S. Silva** (ORCID - 0000-0002-3914-5392) is a Senior Lecturer at the Faculty of Technology, University of Colombo. Her research interests include sensor fusion, assistive technology, embedded systems, neural networks, data analytics and IoT.

**Prasad Wimalaratne** is a Professor at the School of Computing, University of Colombo. His research interests include interactive 3D interfaces, unmanned aerial vehicles (UAVs), virtual environments, assistive technology and code analysis.